\documentclass[preprintnumbers]{revtex4}

\usepackage{graphicx,epsfig,amssymb}
\begin{document}
\bibliographystyle{revtex}

\preprint{RBRC-219}
\preprint{BNL-NT-01/25}
\preprint{YITP-SB-01-63}

\title{Ultrahigh energy neutrinos, small \boldmath{$x$} and unitarity}



\author{Mary Hall Reno}
\email[mary-hall-reno@uiowa.edu]{}
\affiliation{Department of Physics and Astronomy, University of Iowa,
Iowa City, IA 52242 USA}
\author{Ina Sarcevic}
\email[ina@physics.arizona.edu]{}
\affiliation{Department of Physics, University of Arizona, Tucson, AZ 85721 USA}
\author{George Sterman}
\email[sterman@insti.physics.sunysb.edu]{}
\affiliation{C. N. Yang Institute for Theoretical Physics, SUNY Stony Brook,
Stony Brook, NY 11794 USA}
\author{Marco Stratmann}
\email[marco.stratmann@physik.uni-regensburg.de]{}
\affiliation{Institut f{\"u}r Theoretische Physik,
Universit{\"a}t Regensburg, D-93040 Regensburg, Germany}
\author{Werner Vogelsang}
\email[wvogelsang@bnl.gov]{}
\affiliation{Physics Department and RIKEN-BNL Research Center,
Brookhaven National Laboratory,
Upton, NY 11973 USA}



\begin{abstract}
The ultrahigh energy cross section for neutrino interactions with nucleons
is reviewed, and unitarity constraints are discussed. We argue
that existing QCD extrapolations are self-consistent, and do not
imply a breakdown of the perturbative expansion in the
weak coupling.
\end{abstract}

\maketitle

\section{Introduction}

Ultrahigh energy neutrinos are predicted from a number of sources. One
source is from cosmic ray interactions with the microwave background
radiation \cite{Protheroe:1996ft},
producing charged pions which decay into neutrinos. Another possible source
is decaying cosmic strings or extremely massive relics
\cite{Bhattacharjee:2000qc}, which ultimately contribute to a cosmic
neutrino flux. Detection of ultrahigh energy neutrinos
may shed light on the observation of air shower events
with energies in excess of $10^{11}$ GeV, reveal aspects of
grand unification or yield some insight into the sources of the highest energy
cosmic rays.

A number of detectors are able or will be able to detect neutrino induced
showers. For example,
the Auger experiment should be able to detect neutrino induced
horizontal air showers initiated by neutrinos with energies above $10^9$ GeV.
The proposed OWL/EUSO satellite experiments should be able to detect
upward air showers produced by $\nu_\tau\to \tau$ just below
the Earth's surface. The event rates predictions depend on the ultrahigh
energy neutrino cross section, extrapolated beyond the measured regime,
as well as on the predicted neutrino fluxes.

Recent discussions by Dicus, Kretzer, Repko, and Schmidt
\cite{Dicus:2001kb} about the implications of perturbative unitarity
have refocused attention on the ultrahigh energy extrapolation of the
neutrino-nucleon cross section. In the next section we review the
cross section evaluation including the extrapolation of the parton distribution
function to small parton momentum fraction $x$. We examine to what extent
the cross section may be sensitive to the presence of saturation
effects in the evolution of the parton distributions.
In the following section, we outline the unitarity argument
and comment on what can and cannot be learned by relating the neutrino-nucleon
forward scattering amplitude to the total neutrino-nucleon cross section.

\section{Neutrino-nucleon cross section}

The expression for the neutrino-nucleon charged current cross section,
for $\nu_l(k) N(p)\to l(k') X(p')$, is
\begin{equation}
{d^2\sigma\over dx dQ^2}=
{G_F^2\over \pi}\Biggl({M_W^2\over Q^2+M_W^2}\Biggr)^2\cdot
\left[ q(x,Q)+(1-y)^2\,\bar{q}(x,Q)\right]\ ,
\label{dis}
\end{equation}
in terms of $ Q^2=-(k-k')^2$, $x=Q^2/(2M\nu )$ and $y=\nu/E$, for neutrino
energy $E$ and lepton energy transfer $\nu=E-E'$ defined in the nucleon
rest frame. The nucleon mass is $M$, and the center of mass energy squared
is $S=2ME$. In Eq.~(\ref{dis}) we have followed 
Refs.~\cite{Dicus:2001kb,Gandhi:1998ri}
to introduce effective quark and antiquark densities that contain the
contributions from the various flavors as well as the appropriate 
electroweak mixing angles. The expression for the neutral current
reaction $\nu_l(k) N(p)\to \nu_l(k') X(p')$ can be 
cast into an identical form, with the obvious replacement $M_W\to 
M_Z$ and with different effective quark densities. In what follows,
we will only consider the neutrino-nucleon cross section. At high energies,
the antineutrino cross section is expected to be very similar. We 
will neglect perturbative QCD corrections to the cross section which
were found to be small in Refs.~\cite{Dicus:2001kb,Gluck:1999js}.
Finally, we also neglect the contributions to the cross section arising
from charm quarks in the initial state; these can be sizable at
high energies, but are unimportant for our more qualitative purposes.

In Eq.~(\ref{dis}), increasing $Q^2$ has two effects: as $Q^2$ rises, the
cross section decreases due to
the $W$ propagator, but the contributions
of the quark and antiquark distribution functions $q(x,Q)$ and $\bar{q}(x,Q)$
increase due to QCD evolution \cite{Andreev:1979cp}.
The propagator dominates and effectively cuts off
the growth in $Q^2$ at $Q^2\sim M_W^2$. As a consequence, the typical
$x$ value as a function of incident neutrino energy $E$ is
$x\sim {M_W^2 /( 2ME)}$,
so ultrahigh neutrino energies translate to small parton $x$.
For the highest neutrino energies considered, $E\sim 10^{12}$ GeV,
$x\sim 10^{-8}$. HERA measurements of structure functions~\cite{hera}
extend to $x\sim 10^{-6}$; however,
such low values of $x$ are measured at $Q^2<0.1$ GeV$^2$. For
$Q\sim M_W$, the structure functions are measured down to
$x\sim 10^{-3}$ in the D0 experiment's analysis of inclusive jets
\cite{Babukhadia:2001bd}.
Small $x$ extrapolations of the parton distribution functions are therefore
necessary to extend the predictions for the neutrino-nucleon cross section
above $E\sim 10^7$ GeV.

Perturbative QCD governs the small $x$ extrapolations.
The sea quark distributions dominate the cross section
at high energies. Sea quarks are produced by 
gluon splitting $g\to q\bar{q}$, so the gluon distribution
$g(x,Q)$ dictates the eventual quark and antiquark distributions at small $x$.
The gluon distribution is parameterized as $xg(x,Q_0)\sim x^{-\lambda}$
for $x\ll 1$ at a reference scale $Q_0$. Approximate small-$x$ DGLAP
evolution~\cite{dglap}, 
for $\lambda$ close to 0.5, yields a gluon distribution function
of the same form, at a larger value of $Q$: $xg(x,Q)\sim x^{-\lambda}$
\cite{Ellis:1994rb}.
As a practical matter, $\lambda$ was determined at $Q=M_W$ to
extrapolate the parton distribution functions
\cite{Gandhi:1998ri}, for example, those by
the CTEQ collaboration \cite{Lai:1997mg},
below $x=10^{-5}$. Gl\"{u}ck, Kretzer, and Reya \cite{Gluck:1999js}
have checked that the full DGLAP evolution of the Gl\"{u}ck, Reya, and Vogt
\cite{Gluck:1998xa} distribution functions yields only some 20\% difference 
at $x=10^{-8}$ compared with the power law extrapolation of the 
CTEQ densities. Kwiecinski, Martin, and Stasto
\cite{Kwiecinski:1999yf} have performed a BFKL-type~\cite{bfkl} evolution,
yielding results in substantial agreement at the highest energies considered
(10$^{12}$ GeV). The
resulting total neutrino-nucleon cross sections
can be parameterized by  power laws for
$10^7$ GeV$<E<10^{12}$ GeV. For example, the charged current
neutrino-nucleon cross section, using the CTEQ4 parton distribution functions,
scales as $\sigma=5.5\times 10^{-36}(E/{\rm GeV})^{0.36}$ cm$^2$
\cite{Gandhi:1998ri}.

Ultimately, the growth of the parton distribution functions -- and
hence that of the cross section -- predicted by both DGLAP and BFKL 
evolution will have to slow down when the gluon densities become large 
enough that $gg\to g$ recombination processes become 
important~\cite{sat}. The regime in $x,Q^2$ in which this happens 
is referred to as ``saturation region'' and can be estimated. 
It can be roughly characterized by the condition~\cite{gbw}
\begin{equation} \label{satxq2}
Q_s^2 (x) = ({\rm 1 GeV}^2) \cdot \left( \frac{x_0}{x} \right)^{\lambda}
\end{equation}
with $\lambda$ and $x_0$ obtained from a fit to HERA data:
$\lambda=0.288$, $x_0=3.04\times 10^{-4}$. $Q_s$ is the ``saturation
scale'' -- saturation should roughly occur when $Q<Q_s(x)$ at a given $x$.
Clearly, as $x$ decreases, saturation effects are expected to become
relevant already at larger and larger $Q^2$. 

\begin{figure}[htbp]
\vspace*{-1cm}
\centerline{
\includegraphics[width=11.5cm]{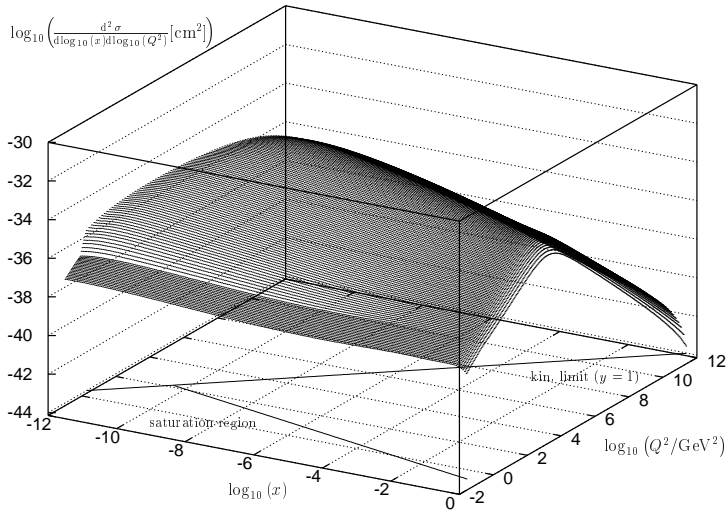}}
\vspace*{-8.3cm}
\caption{The neutrino-nucleon cross section $d^2\sigma/dxdQ^2$ 
at $E=10^{12}$ GeV
as a function of $x,Q^2$. The ``saturation region'' is derived
from Eq.~(\ref{satxq2}). \label{f1}}
\end{figure}

\begin{figure}[htbp]
\vspace*{-1cm}
\centerline{
\includegraphics[width=8.5cm]{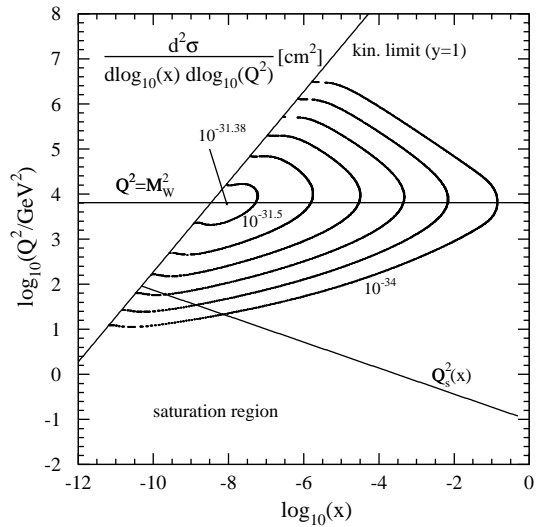}}
\vspace*{-0.4cm}
\caption{Contour plot of the neutrino-nucleon cross section 
in Fig.~\ref{f1} in the $x,Q^2$ plane. \label{f2}}
\end{figure}

In Figs.~\ref{f1} and \ref{f2} we examine whether the neutrino-nucleon 
cross section at ultrahigh energy, $E=10^{12}$ GeV, might be sensitive to 
such saturation effects. Figure~\ref{f1} shows $\log_{10} 
(d^2\sigma/d\log_{10}(x)d\log_{10}(Q^2))$, 
evaluated using the parton distribution 
functions of~\cite{Gluck:1998xa}, as a function of $\log_{10} x$ and 
$\log_{10}Q^2$. It is evident that scales $Q\sim M_{W,Z}$ dominate
the cross section. 
Note that the parton distributions of~\cite{Gluck:1998xa} can be used 
also down to rather small $Q^2\sim 1$ GeV$^2$. This is convenient because,
in order to obtain the total neutrino-nucleon cross section one needs to 
integrate the expression in Eq.~(\ref{dis}) over the range
$0<Q^2<xS$. For $Q<1$ GeV, we have frozen the scale in the parton 
distribution functions to 1~GeV. Fortunately, as can be seen from
Fig.~\ref{f1}, the region of $Q^2<1$ GeV$^2$ contributes only very little
to the total cross section. 

The line in the $x,Q^2$ plane corresponding to the saturation condition, 
Eq.~(\ref{satxq2}), is also shown in Fig.~\ref{f1}. In Fig.~\ref{f2}
we follow Ref.~\cite{Kwiecinski:1999yf} to look at a projection onto 
the $x,Q^2$ plane. The contours are the lines at which the cross
section $d^2\sigma/dxdQ^2$ has fallen to $10^{-31-0.5n}$ cm$^2$,
where $n=1,\ldots ,6$. From these figures, it becomes evident that,
even at the highest neutrino energies, contributions to the cross section
resulting from the regime sensitive to gluon recombination effects are 
marginal. For our example at $E=10^{12}$ GeV, the ``saturation region''
contributes far less than 1\% to the total cross section.

\section{Unitarity considerations}

Dicus, Kretzer, Repko, and Schmidt in Ref.~\cite{Dicus:2001kb} have brought
unitarity considerations to the fore. A restatement of the optical
theorem relates the total neutrino-nucleon cross section to the neutrino-%
nucleon forward elastic scattering amplitude. The latter can
be written in terms of the differential elastic cross section, evaluated
at Mandelstam variable $t=0$, which, with some approximations, yields:
\begin{equation} \label{unit}
{d\sigma_{\rm el}\over dt}\Biggr|_{t=0} \geq {1\over 16\pi}\sigma_{\rm tot}^2
\label{inequal}
\; .
\end{equation}
One can view this as a lower bound on the forward scattering elastic
cross section, or as an upper bound on the total cross section.
In Ref. \cite{Dicus:2001kb}, the authors observe that the inequality
is saturated at a relatively low energy by using the lowest order, $G_F^2$,
contribution for the elastic cross section on the left, and the 
$G_F^4$ contribution
that comes from the inclusive cross section (\ref{dis}) on the right.
Specifically, using the leading term in $G_F$
for the elastic differential cross section, they conclude that
\begin{equation}
\sigma_{\rm tot} \lesssim 9.3\times 10^{-33}\, \mathrm{cm}^2\; ,
\label{size}
\end{equation}
which already is violated
for $E\gtrsim 2\times 10^8\,\mathrm{GeV}$. From 
this they deduce that at yet higher energies, where the right-hand side
of Eq.~(\ref{unit})
increases, while the left is constant (at ${\cal O}(G_F^2)$), previously
neglected terms that are higher order in the weak coupling $g$,
in particular, $g^6$ or $g^8$ terms, {\it must} become important.
They go on to suggest that
this signals a breakdown in perturbation
theory in the weak coupling, $g$.  This is a striking implication indeed,
especially given the small size of the cross section in Eq.~(\ref{size}).

There is another, and we believe more natural, interpretation of
the equality when $E\gtrsim 2\times 10^8\, \mathrm{GeV}$.  
First, we observe that the forward elastic cross section
receives two qualitatively different and quantum mechanically incoherent
contributions.  The first of these describes the coherent elastic scattering
of the entire nucleon through weak vector boson exchange, which begins at
tree level, that is, at $G_F^2$ in the cross section.  The second is
the contribution of high-$Q^2$ virtual states that results from the
{\it incoherent} scattering of partons.   The latter, not the former,
is related independently by the optical theorem to the inelastic cross
section on the right-hand side of Eq.\ (\ref{inequal}), and will saturate
that inequality identically at order $G_F^4$, regardless of its size,
just as at order $G_F^2$ the forward cross section is identically equal
to the corresponding contribution from the square of the real part, 
which has been
neglected on the right of Eq.\ (\ref{inequal}).

That being said, we may still ask whether the dominance of the partonic part 
of the cross section, higher-order by $g^2$
compared to the elastic part, might not be a sign of large contributions
from yet higher orders in the weak coupling. Integrating the factorized 
form Eq.~(\ref{dis}), over $x$ and $Q^2$, however, shows that
at very high energy the square of the total cross section
behaves as $G_F^2$ times $[g^2(S/M_W^2)^{\lambda}]^2$.
This is to be compared to $G_F^2$ on the left-hand side of Eq.~(\ref{unit}).
The factor $g^2$ is the default size of
a higher-order electroweak correction.  The factor 
$(S/M_W^2)^{\lambda}$ is due to the large 
number of partons of size $1/M_W$ at $x\sim M_W^2/S$.
For higher orders in $g^2$ to contribute at a similar level, they would
have to come accompanied by a similar large counting factor.  At the
leading power in $1/M_W$, which is given by Eq.\ (\ref{dis}), this
cannot happen, simply because $q(x)$ and $\bar q(x)$ already count
the partons.  It would still be possible if more partons are involved
in the hard scattering, but this involves going to higher twist,
that is, to explicit suppression by additional powers of $1/M_W$,
which would have to be compensated for by higher-twist multi-parton 
matrix elements.
While such contributions are not very well-known even at low momentum 
transfers, there is no experimental indication of such large scales implicit
within the nucleon.

The forgoing arguments, of course, assume that the unaided QCD extrapolations
described above are equal to the task of so many orders of magnitude.
We have shown above the self-consistency of these extrapolations, and
that they do not, by themselves, lead to problems with unitarity, or give
evidence of a breakdown in perturbation theory in the weak 
coupling~\cite{sigl}.
The very fact of the self-consistency of the QCD extrapolations shows
that ultra high energy neutrinos offer an exploration of the strong
interactions, as well as of cosmic dynamics, into unprecedented
length scales.

\begin{acknowledgments}
We thank Duane Dicus, Stefan Kretzer, and Jamal Jalilian-Marian for 
discussions. This work was supported in part
by NSF Grants No.\ PHY-9802403 (M.H.R.), PHY-0098527 (G.S.),
DOE Contracts DE-FG02-95ER40906 and DE-FG03-93ER40792 (I.S.). 
M.S.\ thanks SUNY Stony Brook and RIKEN, and G.S. and M.S. thank Brookhaven 
National Laboratory for hospitality and support.
W.V.\ is grateful to RIKEN, Brookhaven National Laboratory and the U.S.
Department of Energy (contract number DE-AC02-98CH10886) for
providing the facilities essential for the completion of his work.
\end{acknowledgments}


\end{document}